\documentclass[prx,longbibliography,twocolumn,notitlepage,showpacs,amsmath,amstex,amssymb,citeautoscript,superscriptaddress]{revtex4-1}

\usepackage[english]{babel}
\usepackage{letltxmacro}
\usepackage{latexsym}
\usepackage[shortlabels]{enumitem}

\LetLtxMacro{\ORIGselectlanguage}{\selectlanguage}
\makeatletter
\DeclareRobustCommand{\selectlanguage}[1]{%
  \@ifundefined{alias@\string#1}
    {\ORIGselectlanguage{#1}}
    {\begingroup\edef\x{\endgroup
       \noexpand\ORIGselectlanguage{\@nameuse{alias@#1}}}\x}%
}
\newcommand{\definelanguagealias}[2]{%
  \@namedef{alias@#1}{#2}%
}
\makeatother
\definelanguagealias{en}{english}
\definelanguagealias{English}{english}
\usepackage{graphicx}
\usepackage{amsmath}
\usepackage{mathtools}
\usepackage{amsfonts}
\usepackage{amssymb,bbding}
\usepackage{bm}
\usepackage{color}
\usepackage[percent]{overpic}
\usepackage{soul}
\usepackage{amssymb}
\usepackage{wasysym}
\usepackage{float}
\usepackage{braket}
\usepackage{epstopdf}

\usepackage[normalem]{ulem}

\usepackage{tikz}
\usetikzlibrary{positioning}

\usepackage{blkarray}
\usepackage{multirow}

\usepackage{hyperref}
\hypersetup{
    bookmarks=false,         
    unicode=false,          
    pdftoolbar=false,        
    pdfmenubar=true,        
    pdffitwindow=false,     
    pdfstartview={FitH},    
    pdftitle={},    
    pdfauthor={Authors},     
    pdfsubject={},   
    pdfcreator={},   
    pdfproducer={}, 
    pdfnewwindow=true,      
    colorlinks=true,       
    linkcolor=black,          
    citecolor=blue,        
    filecolor=magenta,      
    urlcolor=blue           
}

\usepackage{soul}
\usepackage{relsize}

\graphicspath{{../imag/}}

\begin{document}

\title{Influence functional of many-body systems: \\ temporal entanglement and matrix-product state representation}

\author{Michael Sonner}\thanks{These two authors contributed equally to this work}
\affiliation{Department of Theoretical Physics, University of Geneva, Quai Ernest-Ansermet 30, 1205 Geneva, Switzerland}
\author{Alessio Lerose}\thanks{These two authors contributed equally to this work}
\affiliation{Department of Theoretical Physics, University of Geneva, Quai Ernest-Ansermet 30, 1205 Geneva, Switzerland}
\author{Dmitry A. Abanin}
\affiliation{Department of Theoretical Physics, University of Geneva, Quai Ernest-Ansermet 30, 1205 Geneva, Switzerland}
\date{\today}
\date{\today}

\begin{abstract}
Feynman-Vernon influence functional (IF) was originally introduced to describe the effect of a quantum environment on the dynamics of an open quantum system. We apply the IF approach to describe quantum many-body dynamics in isolated spin systems, viewing the system as an environment for its local subsystems. While the IF can be computed exactly only in certain many-body models, it generally satisfies a self-consistency equation, provided the system, or an ensemble of systems, are translationally invariant. We view the IF as a fictitious wavefunction in the temporal domain, and approximate it using matrix-product states (MPS). This approach is efficient provided the {\it temporal entanglement} of the IF is sufficiently low.
We illustrate the broad applicability of the IF approach by analyzing several models that exhibit a range of dynamical behaviors, from thermalizing to many-body localized.
In particular, we study the non-equilibrium dynamics in the quantum Ising model in both Floquet and Hamiltonian settings.
We find that temporal entanglement entropy may be significantly lower compared to the spatial entanglement and analyze the IF in the continuous-time limit.
We simulate the thermodynamic-limit evolution of local observables in various regimes, including the relaxation of impurities embedded in an infinite-temperature chain, and the long-lived oscillatory dynamics of the magnetization associated with the confinement of excitations.
Furthermore, by incorporating disorder-averaging into the formalism, we analyze discrete time-crystalline response using the IF of a bond-disordered kicked Ising chain. In this case, we find that the temporal entanglement entropy scales logarithmically with evolution time. The IF approach offers a new lens on many-body non-equilibrium phenomena, both in ergodic and non-ergodic regimes, connecting the theory of open quantum systems theory to quantum statistical physics.

\end{abstract}

\maketitle

\section{Introduction}

The problem of a quantum system interacting with an environment has played an
important role since the early days of quantum mechanics~\cite{BreuerPetruccioneBook}. Describing dynamics of
such {\it open quantum systems} is essential for understanding diverse
phenomena including the process of quantum measurement, transport in nanostructures,
and thermalization~\cite{WeissBook}.

It is an extremely challenging task to describe sufficiently complex, realistic
environments exactly. Therefore much work focused on studying simplified models
of environments, and on developing approximations for open system dynamics. A
model that played a special role is that of an environment made of harmonic
oscillators~\cite{FeynmanVernon} which do not interact among themselves. Interestingly, the dynamics
of a seemingly simple open quantum system -- a two-level system interacting with such a
bosonic environment (or {a bath}) -- exhibits a rich variety of regimes,
depending on the spectral density of the oscillators~\cite{LeggettRMP}.

More recently, experimental advances have enabled realization of many-body
quantum systems which are well isolated from an external environment~\cite{BlochColdAtoms,Blatt12}.
Such setups have brought into focus the problem of highly non-equilibrium dynamics in
{\it closed} many-body systems.
Various universality classes, distinguished by their drastically different dynamical
behavior, have been discovered.

The most common class is that of thermalizing, or ergodic dynamics: starting
from generic initial states, 
local subsystems reach an effectively thermal state under unitary
dynamics~\cite{Polkovnikov-rev}. Remarkably, ergodicity can be
broken by disorder-induced many-body localization (MBL)~\cite{Huse-rev,AbaninRMP,ALET2018498}, as
first foreseen by Anderson in the paper reporting discovery of single-particle
localization~\cite{Anderson58}. Persistence of MBL in periodically driven
(Floquet) systems~\cite{Ponte15,Lazarides15,Abanin20161,Bordia17} has opened the door to
novel non-equilibrium phases such as discrete time
crystals~\cite{Khemani16,Else16} and anomalous Floquet insulators~\cite{Nathan17}. Much
attention has also focused on investigating dynamics in integrable one-dimensional systems, which at long
times relax to a generalized Gibbs ensemble~\cite{EsslerFagottiReview}, and on
weak ergodicity breaking mechanisms such as quantum scars~\cite{ScarsReview}.

At an intuitive level, the character of the system's dynamics is determined by
its properties as a quantum environment. An ergodic system acts as a ``good"
thermal bath for its subsystems, while MBL and other non-ergodic systems fail to
do so. This observation links the dynamics of a many-body system to that of an
open quantum system; however, the environment is now itself a complex
interacting system, rather than a set of non-interacting harmonic oscillators.

The ability of a many-body system to act as a thermal bath has been
characterized using a number of eigenstate and dynamical probes~\cite{AbaninRMP,ALET2018498,Polkovnikov-rev}. The eigenstate
thermalization hypothesis (ETH)~\cite{DeutschETH,SrednickiETH}, as well as its breakdown in MBL systems provide
a particularly useful tool. One may also directly study the dynamics of physical
observables following a quantum quench using numerical techniques such as exact
diagonalization and tensor networks~\cite{VerstrateReview,Schollwock_ReviewMPSTimeEvolution}, and verify whether they settle to thermal
values. The quantum quench setup is routinely employed experimentally in
different platforms, including cold atoms, trapped ions, and superconducting
qubits.

\begin{figure*}[t]
\begin{tabular}{cc}
\centering
\includegraphics[height=0.21\textwidth]{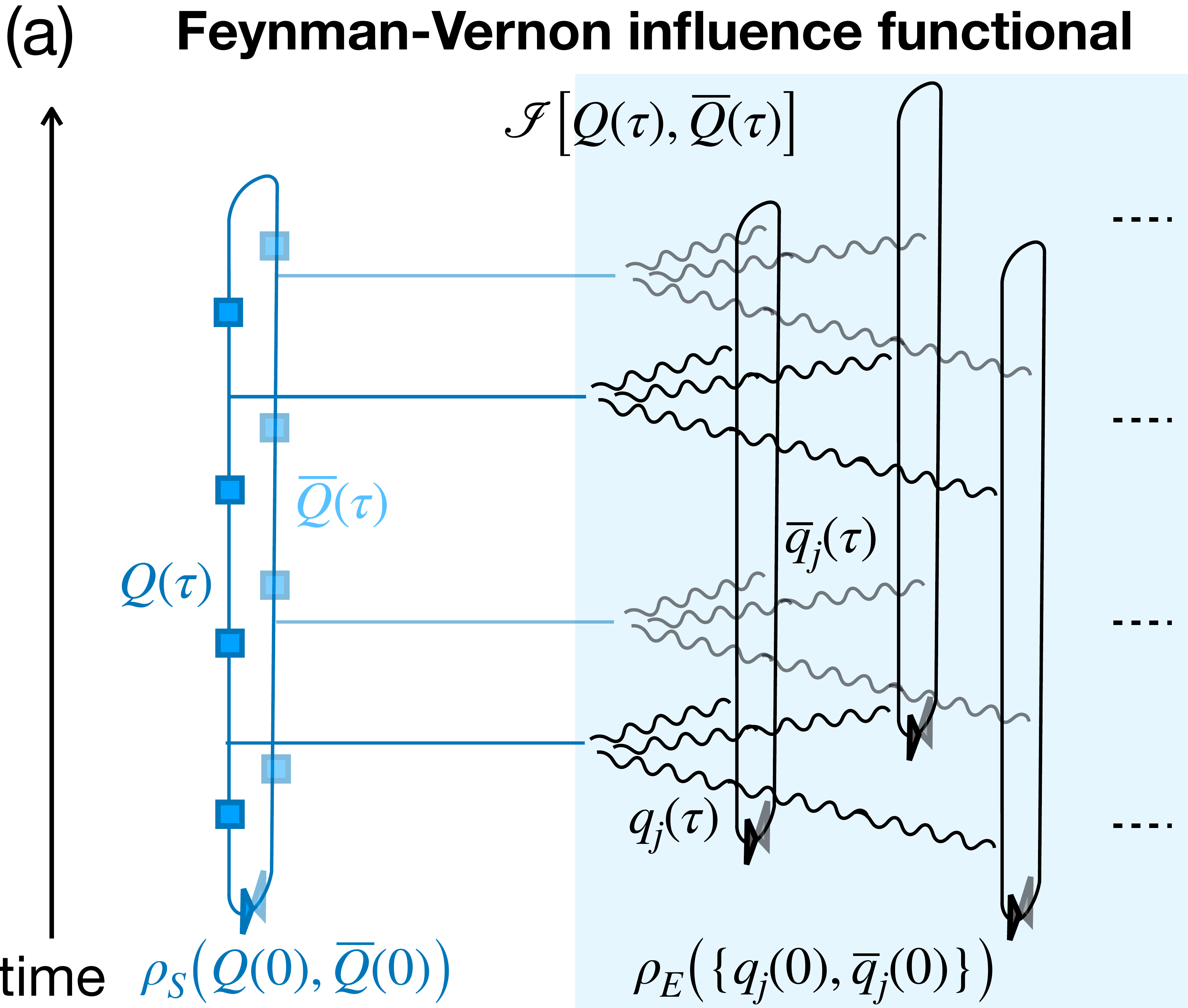}  \hspace{0.25cm}
\includegraphics[height=0.21\textwidth]{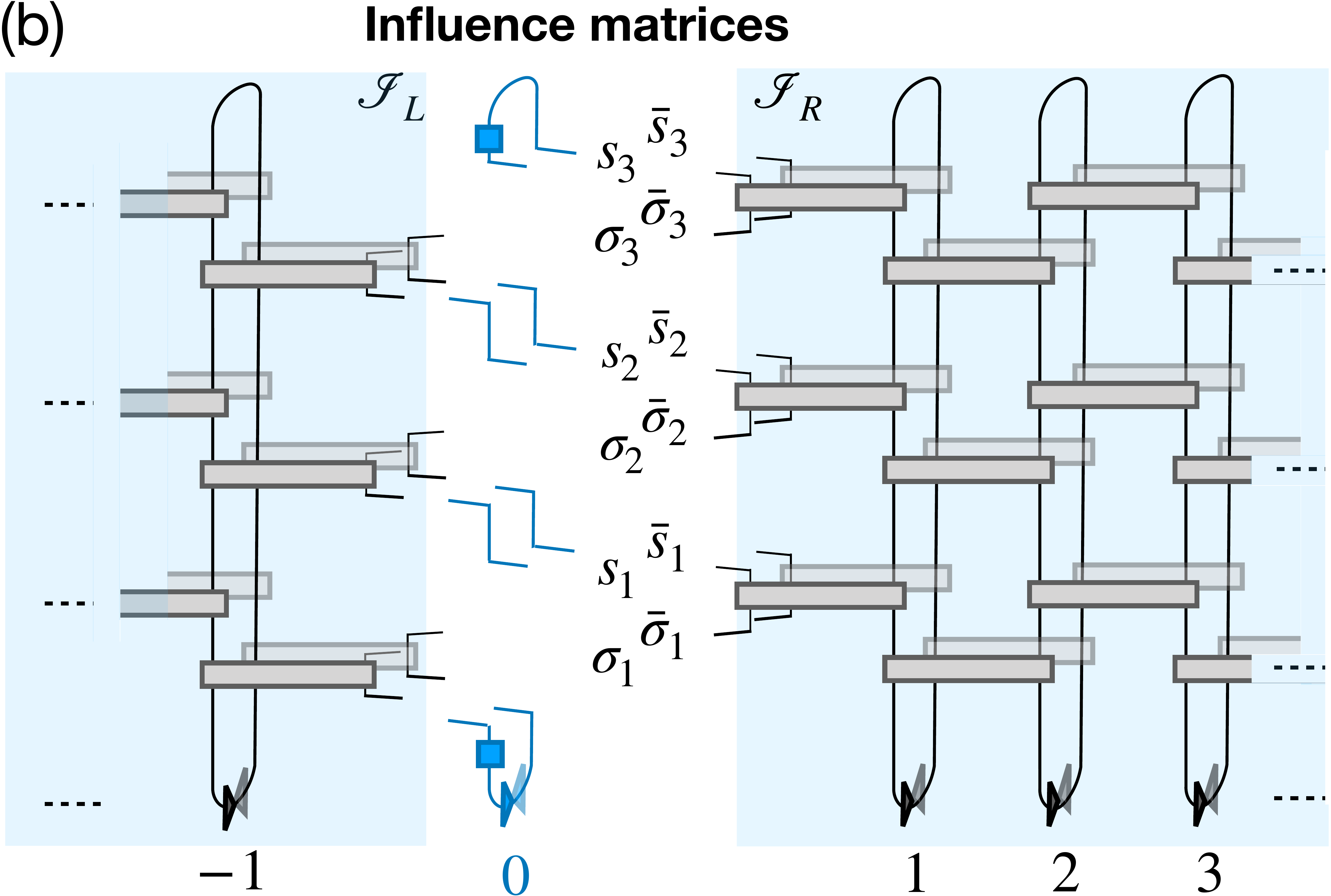}  \hspace{0.25cm}
\includegraphics[height=0.21\textwidth]{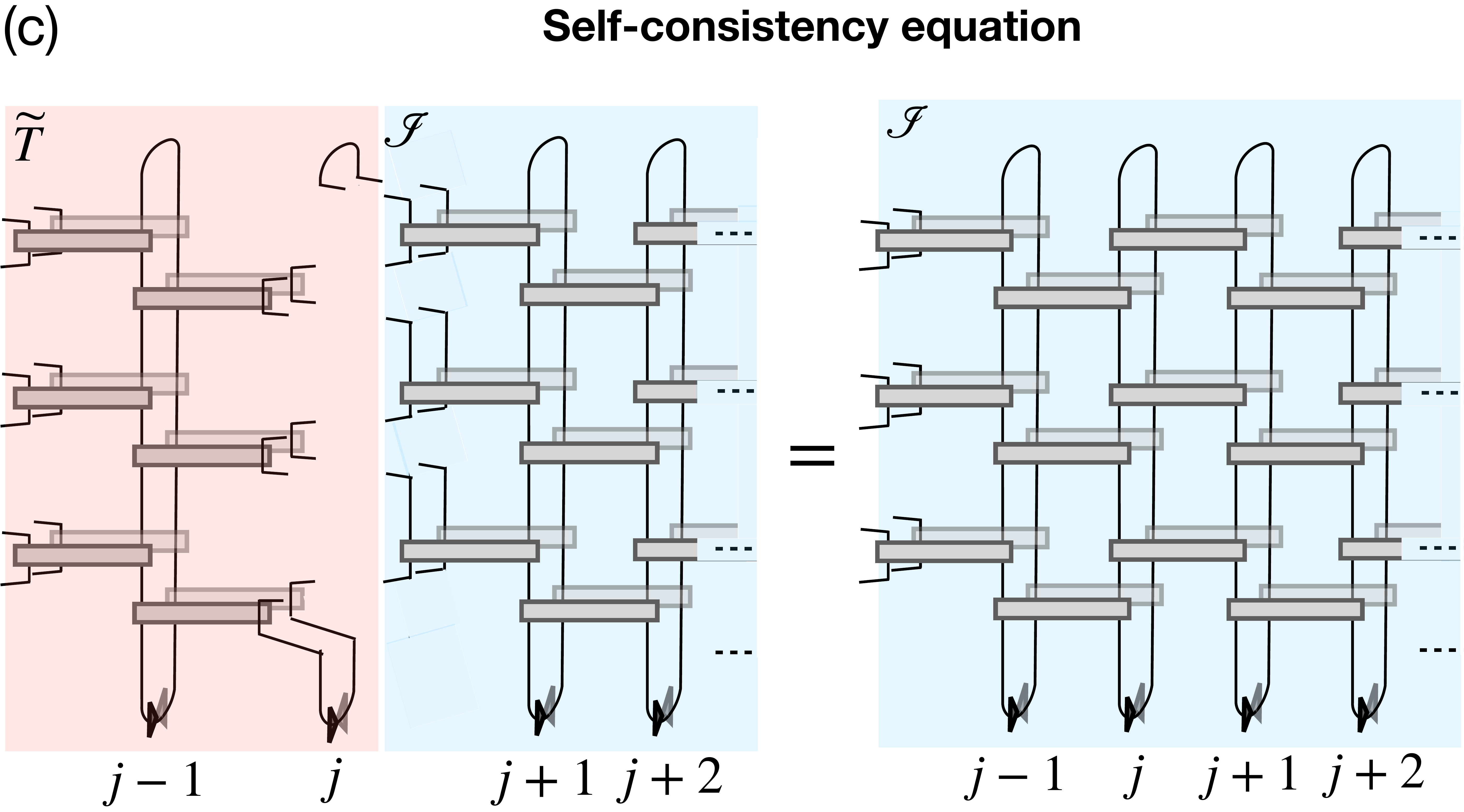}
\end{tabular}
\caption{%
a) Idea of the Feynman-Vernon influence functional for a quantum system interacting with a bath, which consists of non-interacting harmonic oscillators. Considering the time evolution in the Keldysh path integral representation, the bath degrees of freedom $\{q_j(\tau),\bar q_j(\tau)\}$ are integrated out. This yields a contribution $\mathcal{I}[Q(\tau),\bar Q(\tau)]$ in the path integral over the system coordinates only, which is the influence functional.
b) A tensor network describing evolution of a spin chain, in the discrete path integral representation.
Time runs upwards and the foreground (background) represents the forward (backward) branch of time evolution.
Contraction of the parts of the network contained in the blue-shaded boxes defines the left and right influence matrices -- tensors acting on the space of trajectories of spin $j=0$.
For the case of Hamiltonian evolution, when the tensor network represents a discretization of the continuous time evolution, the influence matrices approximate the corresponding continuous-time influence functionals.
c) The influence functional of a homogeneous spin system is an eigenvector of the dual transfer-matrix represented by the red-shaded box; equivalently, it satisfies a linear self-consistency equation (\ref{eq:eigen}).
}
\label{fig_0}
\end{figure*}

In this paper, building on very recent ideas~\cite{LeroseInfluence,Chan21} and earlier related work~\cite{Banuls09}, we employ the
Feynman-Vernon influence functional to analyze the problem of quantum dynamics
and characterize the properties of a many-body system as an environment. The
influence functional (IF) for an open quantum system is obtained by performing
the Keldysh path integral over the environment degrees of freedom
$\{q_j(\tau),\bar{q}_j(\tau)\}$, subject to a time-dependent trajectory of the
system $Q(\tau),\bar{Q}(\tau)$ (here and below each trajectory has a forward and
a backward part, and coordinates without/with a bar parametrize the former/latter).
The idea of this approach is illustrated in Fig.~\ref{fig_0}a: the effect of the
environment is completely described by the IF
$\mathcal{I}[Q(\tau),\bar{Q}(\tau)]$, which weighs the trajectories of the
system. The explicit expression for the IF, available for certain simplified
models such as a bath of harmonic oscillators, has provided a starting point for
describing a two-level system subject to dissipation~\cite{LeggettRMP}.

The approach proposed in Ref.~\cite{LeroseInfluence} is based on the observation
that the IF of a spatially homogeneous many-body system, while difficult to
compute directly, satisfies a self-consistency equation. In the tensor-network
language, this equation amounts to finding an eigenvector of the dual transfer
matrix arising in the path-integral description of the system's dynamics~\cite{Banuls09}, as
schematically illustrated in Fig.~\ref{fig_0}b,c. This eigenvector can be
efficiently sought in the form of a matrix-product state (MPS), provided its
{\it temporal entanglement}  remains low. This is analogous to how MPS-based
algorithms such as the density matrix renormalization group rely on the spatial
entanglement of ground states being low~\cite{VerstrateReview}. However, while the latter is guaranteed
by rigorous results concerning area-law entanglement scaling in gapped systems,
relatively little is known about the scaling of temporal entanglement.

It was appreciated early on~\cite{muller2012tensor} that for certain dynamical problems the approaches
based on 
transverse contractions of the time-evolution tensor network
be computationally more favorable
compared to the 
conventional methods
such as the time-evolving block decimation (TEBD). The recent work~\cite{LeroseInfluence},
which considered periodically driven (Floquet) systems, found that in a family
of thermalizing models, including so-called dual-unitary circuits~\cite{Guhr1,Bertini2019,Piroli2020}, the discrete-time IF -- or influence matrix (IM) -- can be
found exactly and has vanishing temporal entanglement. Furthermore, Ref.~\cite{Koblas20} found a different exact MPS solution for an integrable quantum cellular automaton. The IF approach to a Floquet-MBL model was shown to be efficient as well~\cite{Sonner20CharacterizingMBL}. In a different direction, Refs.~\cite{TEMPO,Cygorek21,Chan21} proposed a tensor-network-based compression of the
IF of baths of noninteracting particles.

In this paper, we extend the IF approach and demonstrate its utility for
several classes of non-equilibrium problems of current interest, both in ergodic and non-ergodic regimes.
We first employ it to describe thermalization dynamics in Floquet and Hamitonian many-body systems. Focusing on spin-$1/2$
chains, we apply the IF formalism to compute the behavior of dynamical
correlators in an infinite-temperature ensemble. Furthermore, we analyze the time-dependent observables
following a quantum quench starting from simple pure initial states. In particular, we simulate the long-lived oscillations of local observables that arise due to confinement of excitations in the quantum Ising chain in a tilted field~\cite{Kormos17,RobinsonNonthermalStatesShort,MazzaTransport,LeroseQuasilocalized}.

In a different direction, below we adapt the IF approach to analyze the dynamics
of non-equilibrium, non-ergodic phases. We focus on a one-dimensional discrete
time-crystal (DTC) enabled by disorder-induced MBL. Our approach allows us to
perform exact averaging over the disordered spin-spin couplings (see below).
We then consider the self-consistency equation for the disordered-averaged IF,
and solve it using an MPS representation.

In this paper we will be particularly interested in the scaling of temporal entanglement (TE)
entropy with the evolution time. We find that in all problems considered, TE remains relatively low up
to sizable evolution times, which indicates efficiency of the MPS
representation of the IF. Furthermore, we investigate the behavior of TE in
Hamiltonian systems as the continuum limit is taken by decreasing the time step
in the Trotterized evolution operator. Surprisingly, TE entropy decreases to
zero in the limit of vanishing time steps; however, as we argue below, small singular values may not always be truncated to faithfully approximate the IF of a Hamiltonian system.

The rest of the paper is organized as follows. In the subsequent Section~\ref{sec2}, we
introduce the IF approach for both Floquet and Hamiltonian dynamics. We also
discuss the MPS representation and the method we use to numerically compute the IF. Subsequently, in Section~\ref{sec3}
we analyze thermalization dynamics in a (non-integrable) quantum Ising
model, presenting results for quenches from different initial
ensembles/states. Both Floquet and Hamiltonian dynamics will be considered. Further, in Section~\ref{sec4} we turn to non-ergodic dynamics and discuss how to incorporate disorder-averaging into the IF formalism, considering discrete time crystal (DTC) as an application. We conclude in Section~\ref{sec5} by discussing future directions and potential of the IF approach to quantum many-body dynamics and
open quantum systems.

\section{Influence functional for many-body dynamics}\label{sec2}

We start our analysis by formulating the influence functional approach to
describing dynamics of a many-body system. Throughout the paper, we focus on
one-dimensional  lattice systems of spins with local Hilbert space dimension $q$ subject to local two-body interactions. We note that the extensions to bosons/fermions, longer-range $k$-body interactions, and to higher-dimensional
systems are straightforward. We consider models in which the time evolution can
be represented by a general ``brick-wall" quantum circuit (Fig.~\ref{fig_0}b). This
class includes Floquet models, and trotterized Hamiltonian evolution. We note that Ref.~\cite{LeroseInfluence} introduced the IF approach for a narrower class of Floquet models.

The central idea of the approach is summarized in Fig.~\ref{fig_0}b,c. The
evolution operator over one period of the brick-wall circuit is given by:
\begin{equation}\label{eq:evolution}
U=U_e U_o, \;\; U_o=\prod_j U_{2j-1,2j}, \;\; U_e=\prod_j U_{2j,2j+1},
\end{equation}
where the two operators correspond to odd and even layers of the quantum circuit.
The two-body evolution operators (unitary gates) in the above expression define the
model in the case of Floquet dynamics. For the case of Hamiltonian evolution,
generated by a two-body local Hamiltonian,
\begin{equation}\label{eq:ham}
H=\sum_i H_{i,i+1},
\end{equation}
the gates are obtained by trotterizing the time evolution, with a time step $\epsilon\to 0$:
\begin{equation}\label{eq:trot}
U_{i,i+1}=e^{-iH_{i,i+1}\epsilon}.
\end{equation}
The physical evolution time $t$ in the Hamiltonian case is the product of the time step $\epsilon$ and
the number of time steps $T$ in the tensor network in Fig.~\ref{fig_0}b, i.e.,
$t=T\epsilon$. We note that the brick-wall quantum circuit can also be used to encode higher-order Trotter schemes; here for simplicity we will consider only the lowest-order scheme.

The time-evolved density matrix of the system after $T$ steps of evolution is
related to the initial state density matrix $\rho^0$ as follows:
\begin{equation}\label{eq:DM}
\rho^{T}=U^{T} \rho^0 {U^\dagger}^T.
\end{equation}
For simplicity, we will consider initial density matrices which can be represented as a tensor product of individual spin density matrices, i.e.,
\begin{equation}\label{eq:DM_prod}
\rho^0=\bigotimes_j \rho_j^0 \, .
\end{equation}
Rewritten via a discrete path integral over intermediate configurations of the
system, the time evolution in Eq.~(\ref{eq:DM}) can be represented by a tensor
network, as illustrated in Fig.~\ref{fig_0}b. We denote the forward and backward trajectory
of spin $j$ by $\sigma_j^\tau,\bar\sigma_j^\tau$, $s_j^\tau,\bar s_j^\tau$, as
shown in the Figure. The gates acting in the forward part of the network
are $U_{j,j+1}$ and the ones in the backward part of the network are
$U^*_{j,j+1}$.

We will be interested in describing the evolution of a sufficiently small
subsystem, e.g. a single spin $j=0$. The dynamics can be characterized by
correlation functions of the form
\begin{align}
\langle \hat{O}_0(T) \hat{O}_0(0) \rangle,
\label{eq:localcor}
\end{align}
where $\hat{O}_0$ is an operator representing a local observable at site $j=0$,
see Fig.~\ref{fig_0}b.
To calculate such correlators we begin by tracing out the environment degrees of
freedom, i.e., the other spins; the spins with $j>0$ can be viewed as a ``right"
environment, and the ones with $j<0$ as a ``left" environment. This procedure yields right and left influence matrices
$\mathcal{I}_{R(L)}$, which are formally introduced by contracting tensor
networks contained in the blue-shaded boxes in Fig.~\ref{fig_0}b, with a
boundary defined by spin $j=0$ trajectory; in this contraction, the summation
over all final states of the spins in the environment is carried out. We can
now calculate any correlator of the form Eq.~\eqref{eq:localcor} by contracting
the IMs with a tensor network for evolution of spin $j=0$, with operators $\hat{O}_0$ inserted at $\tau=0,\, T$.
In the Hamiltonian case, the IMs approximate the corresponding continuous-time IFs as $\epsilon\to0$.
(In the following we will often use the names IM and IF interchangeably.)

In certain cases, which include dual-unitary circuits/perfect
dephasers~\cite{Guhr1,Bertini2019,Piroli2020,LeroseInfluence} and certain integrable
systems 
~\cite{Koblas20,IntegrablePaper}, the
contraction of the tensor network that defines the IF can be carried out
analytically. Generally, the IF for an environment with two additional spins can
be obtained from the previous IF by applying one vertical slice of the tensor
network in Fig.~\ref{fig_0}c (red shaded box). These vertical slices define the
{\it dual transfer matrices}~\cite{Banuls09}, which act on the $q^{4T}$-dimensional space of
single-spin trajectories. As discussed below, for certain systems contracting the tensor
network in the space direction using dual transfer matrices proves advantageous
compared to the standard contraction in time direction.

In translationally invariant systems, the dual transfer matrix
$\tilde{T}$ is the same at every step. Notice that the strict light
cone in our model means that the IF that describes evolution over time $T$ can depend at most on the motion of
closest $2T$ spins. Hence the IF of a homogeneous system is a unique
eigenvector of $\tilde{T}$ with an eigenvalue~$1$. The eigenvalue
equation for the right IF
\begin{equation}\label{eq:eigen}
\tilde{\mathcal{T}} |\mathcal{I}\rangle=|\mathcal{I}\rangle,
\end{equation}
(where we omitted the subscript  $R$ for simplicity) has a simple physical
interpretation: it can be viewed as a {\it self-consistency equation} for the
IF. In essence, a spin evolving under the influence of the right environment, exerts
the same influence on its left neighbor. The left influence functional clearly
satisfies an analogous equation with a slightly modified transfer matrix.

We note that this approach can be extended to describe ensembles of disordered
models with spatially uncorrelated, translationally-invariant distribution: in
that case, the transfer matrix in Eq.~(\ref{eq:eigen}) is disorder-averaged, and
its eigenvector corresponds to an IF of a disorder-averaged environment.
In Ref.~\cite{Sonner20CharacterizingMBL}, a particular Floquet-MBL model with on-site disorder
fields was considered (see also Refs.~\cite{SchuchMBL,chan2020spectral,garratt2020manybody}). Below in Section IV we will discuss a generalization of the method
to a model with disordered two-body interactions, which exhibits discrete time crystal behavior.

A promising approach to solving the self-consistency equation (\ref{eq:eigen})
is based on approximating the 
influence matrix 
by an MPS with a fixed bond dimension~\cite{Banuls09,muller2012tensor,LeroseInfluence},
similarly to conventional techniques for two-dimensional systems~\cite{HaegemanReview}. 
This approach is expected to be efficient when TE
entropy is 
low. In that case, the IF formulation may allow one to analyze
time-dependent correlation functions which are out of reach of standard methods
such as TEBD, exact diagonalization, etc.

Interestingly, temporal entanglement can be low in different dynamical regimes.
A striking example is provided by perfect-dephaser Floquet circuits: for an
infinite-temperature initial density matrix and certain pure initial states in MPS form, the IF is a product state with zero
TE~\cite{LeroseInfluence,Piroli2020}; in this case, the environment effectively {dephases}
local spins at every step of Floquet evolution.  TE remains low in the vicinity
of such solvable points, which ensures the efficiency of the MPS representation.
Furthermore, the TE was found to be low up to long times in certain
non-integrable Hamiltonian
models~\cite{muller2012tensor,HastingsPRA14}, and in proximity to integrability~\cite{IntegrablePaper}. Ref.~\cite{Sonner20CharacterizingMBL}
further demonstrated that the approach remains efficient in the ergodicity-breaking
Floquet-MBL phase, thanks to slow, sub-linear scaling of TE of the
disorder-averaged IF. Despite these results, the full potential of the IF-based
approach to describe quantum dynamics remains to be investigated; the key open
question concerns the scaling of TE with the evolution time and the formulation of suitable numerical algorithms to approximate it.

In what follows, we will analyze several Hamiltonian and Floquet models, both in the ergodic and
non-ergodic regimes. We will compute dynamical correlation functions using the IF approach
for both mixed and pure initial states of the system. We will also investigate
the scaling of TE, and its dependence on the system's parameters, and on the
time evolution step $\epsilon$, for the case of Hamiltonian dynamics.

\section{Thermalizing dynamics in Floquet and Hamiltonian systems}\label{sec3}

In this Section, we will study thermalization dynamics in non-disordered Floquet and Hamiltonian models in different setups using the IF approach.
 We will consider relaxation of observables in a chaotic system initialized in an infinite-temperature ensemble, both for a spin that is part of the system, and for an impurity embedded in a many-body environment. Further, quench dynamics from a pure initial state in a model with confined excitations will be studied. For all examples considered, we will investigate the behavior of TE entropy with evolution time.

\subsection{Floquet dynamics}

\begin{figure*}[t]
\centering
\includegraphics[width=0.7\textwidth]{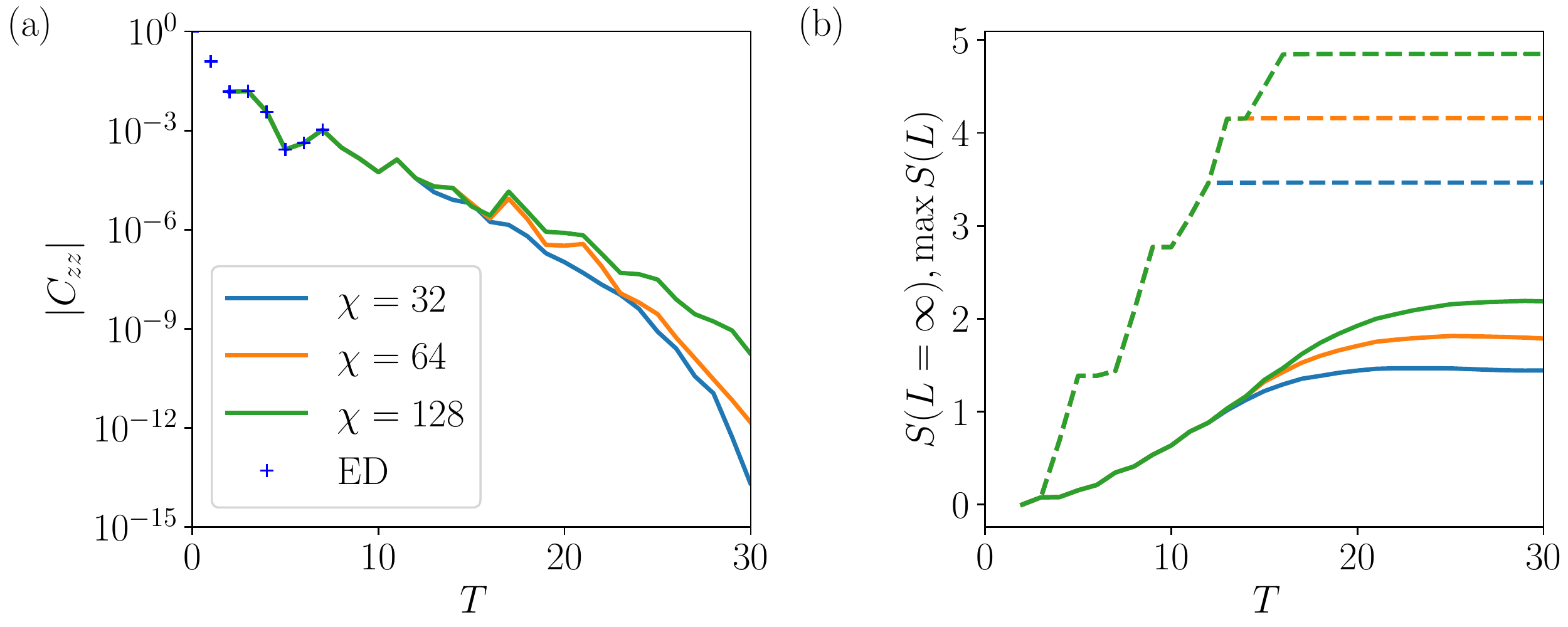}
\caption{Simulation of the kicked Ising model Eq.\eqref{eq:KIM} with parameters
$J=0.8, g=0.7236,h=0.6472$ \cite{Kim_ETH} for an infinite-temperature initial state.
{\it Left:} Autocorrelation function of the $\sigma^z$ operator as a function of
time. Exponential relaxation signals chaotic dynamics. The blue crosses
correspond to exact diagonalization results, which coincide with the MPS calculations. The results
are well converged up to time $T\approx 20$. {\it Right:} Temporal entanglement
entropy as a function of evolution time for different bond dimensions. Dashed lines are
maximal entanglement entropies which are encountered during iteration process starting
from open boundary IF. Starting from the PD IF, the entropy of the final IF is
also the maximal entropy (solid lines). This illustrates the importance of the choice of the
initial IF in the context of the iterative MPS method.}
\label{fig2}
\end{figure*}

We start by illustrating the use of the method for Floquet systems. As an
example of a thermalizing Floquet model, we choose the kicked Ising model for
spins-$1/2$, which has been extensively
studied in the literature~\cite{Kim_ETH,Guhr1,Bertini2019}. The Floquet operator of this model is
given by:
\begin{equation}\label{eq:KIM}
U=e^{-ig \sum_j \sigma_j^x} \, e^{-iJ\sum_j \sigma_j^z \sigma_{j+1}^z -ih\sum_j \sigma_j^z},
\end{equation}
where $\sigma_j^\alpha$, $\alpha=x,y,z$ are standard spin-$1/2$ Pauli operators.
Depending on the values of parameters $J,g,h$, the model displays a rich variety
of dynamical behavior. For $h=0$, it can be mapped onto a free-fermion
model using a Jordan-Wigner transformation, and is therefore integrable. In this
case, it can be shown that the IF obeys an area-law temporal entanglement, even
in the limit of very long evolution times, $T\to\infty$~\cite{IntegrablePaper}.

Furthermore, the model has special properties at the self-dual points
$|J|=|g|=\pi/4$~\cite{Guhr1,Bertini2019}. In that case, even though dynamics at
$h\neq 0$ are chaotic, many quantities can be computed analytically, including
various dynamical correlation functions. At these points, the IF has a
perfect-dephaser form~\cite{LeroseInfluence}. When parameters of the model are detuned slightly away from these PD
points, the temporal entanglement remains low, showing an apparent volume-law
scaling with $T$, albeit with a small prefactor; thus, the IF can be
well-approximated by a MPS with a moderate bond
dimension~\cite{LeroseInfluence}.

Here, we focus on the parameter choice $J=0.8, g=0.7236,h=0.6472$, considered in
Ref.~\cite{Kim_ETH}. It was found that such a model displays chaotic dynamics, and,
moreover, the deviations from the eigenstate thermalization hypothesis
quickly become negligible already at moderate finite sizes. We computed the IF
for the infinite-temperature initial density matrix,
\begin{equation}
\rho_j^0=\frac{1}2  \left( \begin{array}{cc}
1 & 0 \\
0 & 1
\end{array}\right)
\end{equation}
by iteratively applying the dual transfer matrix to an initial ``boundary" IF. As discussed below, certain choices of the boundary IF can significantly improve the performance of the MPS approach.

To perform the tensor network contraction, we set up a code using the tenpy library~\cite{tenpy}.
The diagonal form of interactions in the $z$-basis allows us to identify the input and output legs at each
time step (denoted by $s$ and $\sigma$ in Fig.~\ref{fig_0}). Thus, compared to the general circuits, the number of
degrees of freedom in the dual transfer matrix and in the IM is reduced by a factor of $2$ ~\cite{LeroseInfluence}; moreover, 
the maximum velocity of propagation is $1$ rather than $2$. The absence of correlations in the initial state of the chain combined with the strict light cone ensures that  the
IF reaches its self-consistent thermodynamic limit form after at most $T$
iterations of the dual transfer matrix $\tilde{T}$. The latter can be
straightforwardly represented as an MPO of bond dimension $4$ for the model (\ref{eq:KIM}). We select a maximum
bond dimension $\chi$ and iteratively apply this MPO to a boundary vector
expressed in an MPS form, truncating the result to $\chi$ using SVD compression
when necessary.

The choice of the boundary IF does not affect the final IF obtained after $T$ iterations, provided the calculations are carried out exactly. However, different choices generally lead to different intermediate IFs encountered during iterations. Physically, the $\ell$-th intermediate IF describes the effect of an environment that consists of $\ell$ spins, which evolve under unitary evolution set by the brick-wall quantum circuit, and are subject to an external bath at the boundary; the properties of this bath are determined by the chosen
initial IF. In order for MPS-based methods to be efficient, it is important
that the TE of the intermediate IF remains low enough, such that truncation does not incur a significant error. However, this is not always the
case. Interestingly, seemingly the most natural choice of the boundary IF that corresponds to open boundary
conditions, $\mathcal{I}[\sigma_\tau,\bar\sigma_\tau ] = 1$, turns out not to be the most efficient one: in this case TE grows as a function of $\ell$ first before decreasing again to the final value. We found that the maximal TE entropy of
intermediate IFs can be very high; its growth as a function of evolution time is shown by the dashed lines in
Fig.~\ref{fig2}b.  We note that a similar observation was reported in Ref.~\cite{Chan21} for different models.

However, the problem of the ``entanglement barrier" described above can be circumvented by choosing a different boundary IF. For the thermalizing Floquet model, we chose the perfect dephaser IF
$I[\sigma_\tau,\bar\sigma_\tau] =\prod \delta_{\sigma_\tau,\bar\sigma_\tau}$ as a
boundary condition. Physically, this corresponds to a boundary where the spin is
measured (in the $\sigma^z$-basis) every period. We confirmed that this choice indeed leads to a monotonic
growth of entanglement as a function of iteration number $\ell$, and therefore the ``entanglement barrier" can be avoided.


%

Using the MPS representation of the IF we computed the dynamical correlation
function:
\begin{equation}
\label{eq_czzdef}
\mathcal{C}_{zz}(t) =
\Big\langle
\hat \sigma^z_0 (t) \hat\sigma^z_0(0)
\Big\rangle \, .
\end{equation}
The resulting exponential decay of this correlator, illustrated in
Fig.~\ref{fig2}a, is consistent with the scrambling of local quantum information
expected in this chaotic system.

Fig.~\ref{fig2}b shows the scaling of TE entropy (for a bipartition at $\tau=T/2$) with
the evolution time computed with the MPS approach for three choices of the maximum
bond dimension. The convergence of TE for different bond dimensions up to
$T\approx 20$ indicates that the MPS form provides a good approximation to the exact IF.
For times up to $T=7$ this was verified by comparing the results to exact ones found
with the standard sparse linear algebra library {ARPACK} (blue crosses in Fig.~\ref{fig2}). TE entropy follows an apparent volume-law
scaling in the accessible time interval. We note that at present it is unclear
whether this scaling persists as $T \to\infty$.

\subsection{Hamiltonian dynamics}





\begin{figure*}[t]
\centering
\includegraphics[width=\textwidth]{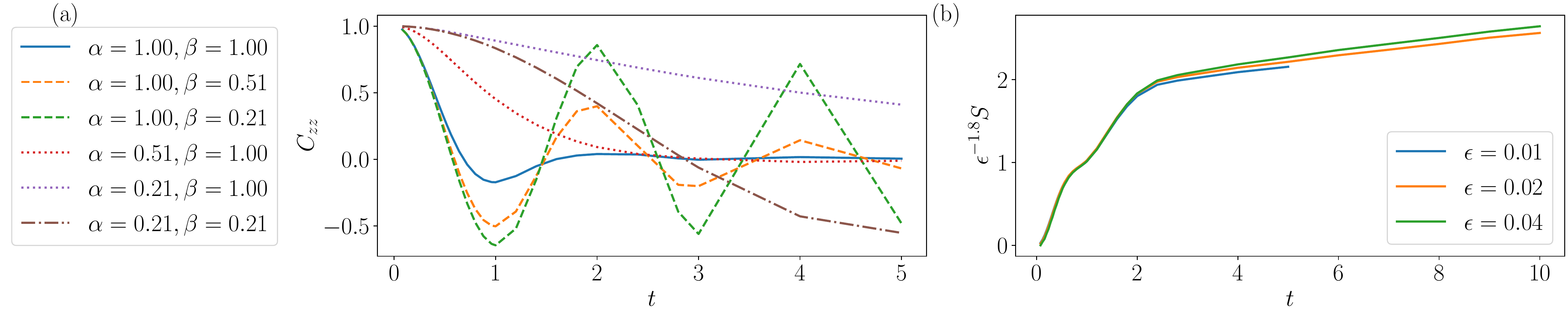}
\caption{{\it Left:} Autocorrelation $\mathcal{C}_{zz}$ of Hamiltonian model
\eqref{eq_qic} for parameters $J=1.0,g=\sqrt{2},h=0.681$ at infinite temperature
with different impurities. The parameters were chosen to be incomensurate to
avoid any resonances which may affect dynamical properties. The result for a homogeneous chain is plotted by the solid
line. We can observe the perturbative (Born) limit by tuning the coupling of the impurity
spin to the spin chain (dotted lines) and the memoryless (Markovian) limit by tuning the driving
of the impurity (dashed lines) as well as the combined Born-Markov limit
(dash-dotted line). The calculation is fully converged for bond dimensions
$\chi=64,128$ and time steps $\epsilon=0.01,0.02,0.04$.  {\it Right:} TE entropy scales
to zero as a function of the time step $\epsilon$, in a manner that is approximately consistent with a power-law dependence on $\epsilon$ in the parameter range considered.}
\label{fig3}
\end{figure*}

We can use a related approach to study Hamiltonian dynamics, by viewing
Eq.~\eqref{eq:KIM} as a (second-order) trotterization of the continuous time
unitary flow $U=e^{-iHt}$ generated by the Hamiltonian
\begin{equation}
\label{eq_qic}
H= -J\sum_j \sigma_j^z \sigma_{j+1}^z -h\sum_j \sigma_j^z -g \sum_j \sigma_j^x.
\end{equation}
This Hamiltonian describes a quantum Ising chain in a tilted magnetic field. The
identification is realized by rescaling the parameters in the Hamiltonian
Eq.~\eqref{eq_qic} with the time step $0 < \epsilon\ll 1$ in order to obtain the
dimensionless parameters  $ J,g,h \mapsto J \epsilon,g
\epsilon,h \epsilon$ in Eq.~\eqref{eq:KIM}; in this case, the physical time is $t=T\epsilon$.
Higher-order Trotter schemes can be implemented analogously.

The model in Eq.~\eqref{eq_qic}, despite its apparent simplicity, exhibits a rich variety of phenomena, and therefore it provides a fruitful playground in quantum many-body physics, both in and out-of-equilibrium. Similarly to its Floquet counterpart, it is integrable when
the magnetic field is purely transverse ($h=0$), and exhibits a paradigmatic
quantum phase transition between a paramagnetic $|g|>J$ and a ferromagnetic
$|g|<J$ phase, corresponding to a spontaneous breaking of the $\mathbb{Z}_2$
symmetry.
The free fermionic quasiparticles of the model change their nature across the
transition, from non-topological, spin-wave-like to topological,
domain-wall-like excitations interpolating between the two degenerate
vacua~\cite{Sachdevbook}. In the latter case, the longitudinal field $h$
explicitly breaks the symmetry, lifting the degeneracy of the two magnetized
ground states and inducing a first-order transition. Accordingly, a confining
potential $V(r) \propto h r$ arises between a pair of domain walls, which bind
together into ``mesonic'' composite particles~\cite{McCoyWuConfinement}. The
model with nonvanishing fields $g,h\neq0$ is non-integrable and its dynamics are
believed to be generally chaotic~\cite{Kim_ETH}, although anomalously slow or
even suppressed relaxation after a quench has been reported even far away from
solvable limits~\cite{BanulsWeakThermalization,LinMotrunichOscillations,Wurtz20}
and related to confinement of
excitations~\cite{Kormos17,RobinsonNonthermalStatesShort,MazzaTransport,LeroseQuasilocalized}.

First, we study the local relaxation of a slower impurity spin at position $j=0$ embedded in the quantum Ising chain at infinite temperature. This impurity is characterized by having smaller values of the fields $g_0= \alpha g$, $h_0=\alpha h$ and of the couplings $J_{-1,0}=J_{0,1}=\beta J$ to the neighboring spins.
Note that the IF only depends on $\beta$ and not on $\alpha$.
Tuning $\beta$ and $\alpha$ to be much smaller than $1$, the impurity dynamics approaches a typical open quantum system (OQS) setup, i.e., that of a two-level system  weakly coupled to a much faster environment, which allows for the celebrated Born (for $\beta\ll 1$) and Markov (for $\alpha\ll 1$) approximations, respectively. Putting $\beta$ and $\alpha$ to $1$, instead, we retrieve the standard homogeneous quantum many-body system setup, where the distinction between the relevant subsystem and its bath is purely conceptual.
Fig.~\ref{fig3} illustrates representative results of these computations.
The left panel shows the infinite-temperature time-dependent autocorrelator $\mathcal{C}_{zz}(t)$ [cf. Eq.~\eqref{eq_czzdef}]
converged with respect to both Trotter step and bond dimension, for decreasing $\alpha$ and $\beta$ from the homogeneous case to an OQS limit.
The right panel shows the scaling of TE entropy $S$ converged  with respect to the bond dimension  vs the physical evolution time $t$ (i.e., for $T=t/\epsilon$ steps), for a range of decreasing Trotter steps $\epsilon$.
Interestingly, for fixed $t$, the value of $S$ scales to \textit{zero} for $\epsilon\to0$ with an apparent power law with an exponent $\simeq1.8$. We will further comment on this observation below.

\begin{figure*}[t]
\centering
\includegraphics[width=0.9\textwidth]{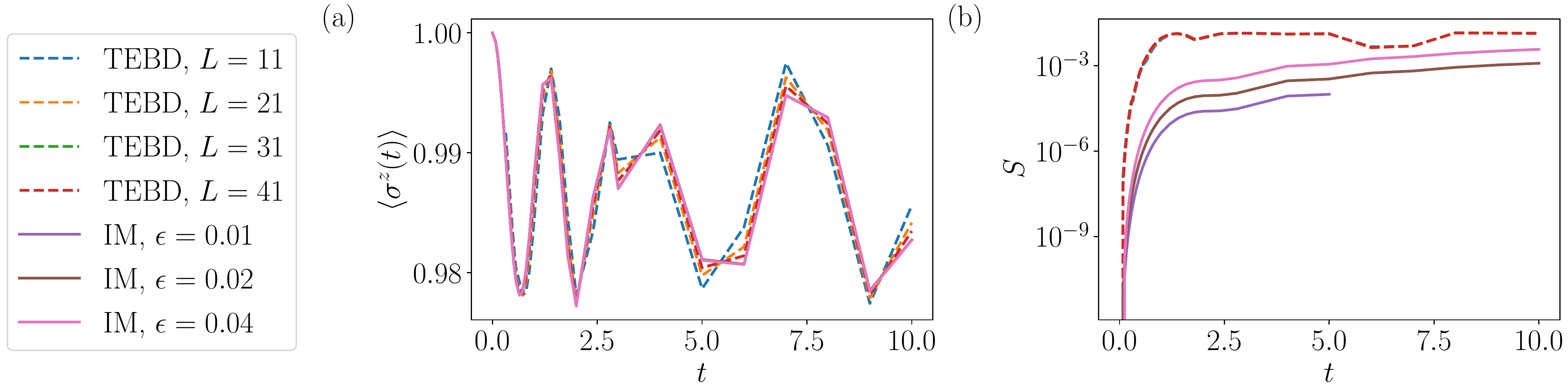}
\caption{Comparison of TEBD and IM methods for the Hamiltonian model \eqref{eq_qic} at
$J=1.0,g=0.25,h=0.4$ undergoing a quench from the polarized state (cf Refs.~\cite{BanulsWeakThermalization,Kormos17}). In both cases, bond dimension $\chi=64$ was used. {\it Left:} While both curves are converged with respect to
time step and bond dimension, the TEBD results still show small finite size effects. {\it Right:} The real
space entanglement entropy of the TEBD MPS is significantly higher than the TE entropy for
all three time steps, indicating the lower computational cost of the IM approach.}
\label{fig4}
\end{figure*}

Second, we study a prototypical quantum quench setup, where the (now
impurity-free) system is initialized in a fully polarized state in the positive
$z$ direction. We focus on the ferromagnetic phase, setting $g=0.25 J$ {and $h=0.4J$}. The
evolution of the ferromagnetic order parameter $\langle \sigma^z_0(t) \rangle$
in Fig.~\ref{fig4} displays clear, long-lived quasiperiodic oscillations,
consistent with the discrete tower of mesonic particles excited by the sudden
quench~\cite{Kormos17}. The right panel shows that the same scaling of the maximum TE entropy with
the Trotter step is found even for quench dynamics, cf. Fig.~\ref{fig3}.
In this case, furthermore, it makes sense to compare spatial entanglement
entropy to temporal entanglement entropy. As the plot clearly shows, the former
-- obtained by TEBD converged with system size and bond dimension -- significantly exceeds the latter, while still remaining relatively low as a consequence of
confinement~\cite{Kormos17,LeroseQuasilocalized}.

\subsection{Continuous-time limit}

The simulations presented in this Section were obtained with a second-order
Trotter scheme with time steps $\epsilon =0.01,0.02,0.04$. The Trotter error was found to be negligible in
all cases for $\epsilon\leq 0.04$: in the examples shown in Figs.~\ref{fig3},~\ref{fig4},
time-dependent observables were perfectly converged with respect to decreasing
$\epsilon$. Thus, effectively, the continuous time limit is well captured by the discretized evolution.

On the methodogical side, it is interesting to discuss the non-trivial scaling of temporal
entanglement when $\epsilon$ is decreased. Surprisingly, as shown in the right panel of Fig.~\ref{fig3}, the maximum temporal
entanglement entropy $S_t$ at fixed physical evolution time $t$ scales to
\textit{zero} for $\epsilon \to 0$. This behavior was found in all our simulations of Hamiltonian dynamics. To gain an insight into the origin of this behavior, below we analyze several solvable limits.

%

First, we study the continuous-time limit of the IM in
the limiting case $g=0$,  where quantum fluctuations are completely suppressed
and all product states in the $\sigma^z$-basis are eigenstates of the Floquet
operator/Hamiltonian for all $J$, $h$. Let us first consider the Floquet problem of
Eq.~\eqref{eq:KIM}, with an infinite-temperature initial ensemble. The analytical form of the IM can be found exactly:
\begin{equation}
\label{eq_IMg0}
I[\sigma_\tau,\bar\sigma_\tau] = \cos \Big[ J   \sum_\tau (\sigma_\tau - \bar\sigma_\tau ) \Big].
\end{equation}
It is straightforward to verify that this expression satisfies the self-consistency equation for $g=0$.
{Physically, this expression can be thought of as the influence functional due to an interaction with a constant classical magnetic field along $\hat z$-axis, which takes values $+J$ or $-J$ with probability $1/2$.}
To compute the von Neumann entropy of $I$, we need to normalize it as a pure wavefunction in the folded Hilbert space, i.e.,
\begin{widetext}
\begin{equation}
    \ket{\Psi} =
    \frac 1 {\sqrt{2[1+\cos^{2T}(2J)]}}
    \bigg(
    \bigotimes_\tau \frac 12
    \begin{pmatrix}
    1 \\ 1
    \end{pmatrix}_{cl,\tau}
    \otimes
    \begin{pmatrix}
    e^{2iJ} \\ e^{-2iJ}
    \end{pmatrix}_{q,\tau}
    \; + \;
    \bigotimes_\tau \frac 12
    \begin{pmatrix}
    1 \\ 1
    \end{pmatrix}_{cl,\tau}
    \otimes
    \begin{pmatrix}
    e^{-2iJ} \\ e^{+2iJ}
    \end{pmatrix}_{q,\tau}
    \bigg)
\end{equation}
\end{widetext}
where we have factored the four-dimensional folded Hilbert space at time $\tau$
into the tensor product of a ``classical''
$\mathrm{Span}(\ket{\uparrow\uparrow},\ket{\downarrow\downarrow})$ and a
``quantum'' $\mathrm{Span}(\ket{\uparrow\downarrow},\ket{\uparrow\downarrow})$
two-dimensional subsystems.
The chosen prefactor ensures the normalization condition, $\braket{\Psi|\Psi}=1$.

Our aim is to compute the bipartite von Neumann entanglement entropy of this normalized state wavefunction with respect to a bipartition $(1,M),(M+1,T)$.
First, we notice that the classical sector does not contribute to entanglement, since the corresponding degrees of freedom are in a product state.
Second, restricted to the quantum  sector, $\ket{\Psi}$ can be thought of as an equal-weight quantum superposition of two pure spin-coherent states of a spin of magnitude $T/2$,  fully polarized along the directions $\cos(2J) \hat x \pm \sin(2J) \hat y$, respectively.
These two spin-coherent states can be pictured as Gaussian wavepackets on the surface of the Bloch sphere, centered around those two directions, with transverse fluctuations of width equal to $1/\sqrt{T}$ (relative to the unit radius of the Bloch sphere).
Since the two polarization directions form an angle $4J$, it is easy to see that the overlap of the two spin-coherent states decreases exponentially with $J^2 T$. 
For $J^2 T \ll 1$, the two states have large overlap, so $\ket{\Psi}$ is close to a product state, giving a vanishing von Neumann entanglement entropy. In contrast, for $J^2 T \gg 1$, the two states are effectively orthogonal, such that $\ket{\Psi}$ becomes a GHZ state with von Neumann entanglement entropy $\log 2$. The exact calculation gives
the binary entropy
\begin{equation}
    S_{M,T}(J) = - P \log P - (1-P) \log (1-P),
\end{equation}
 {with }
 \begin{equation}
    P = 1 - \frac {\cos^{2M}(2J)+\cos^{2(T-M)}(2J)}{1+\cos^{2T}(2J)}\, ,
\end{equation}
which exhibits the expected features.
Setting $x=J^2 T$, $0\le f=M/T\le 1$, this quantity becomes a smooth function of $x$ and $f$ in the limit $T\to\infty$.

\begin{figure*}[t]
\centering
\includegraphics[width=0.29\textwidth]{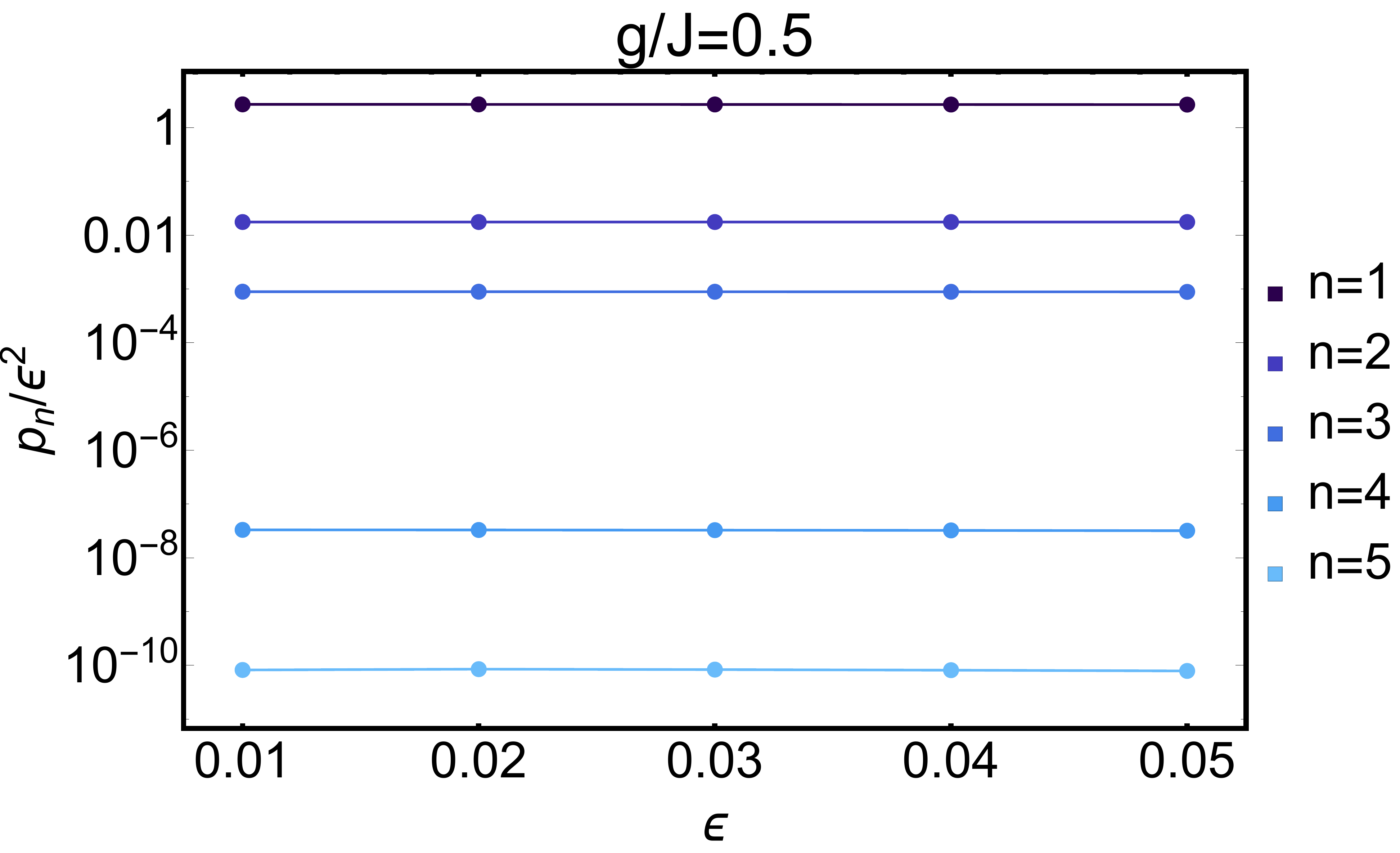}
\hspace{0.25cm}
\includegraphics[width=0.29\textwidth]{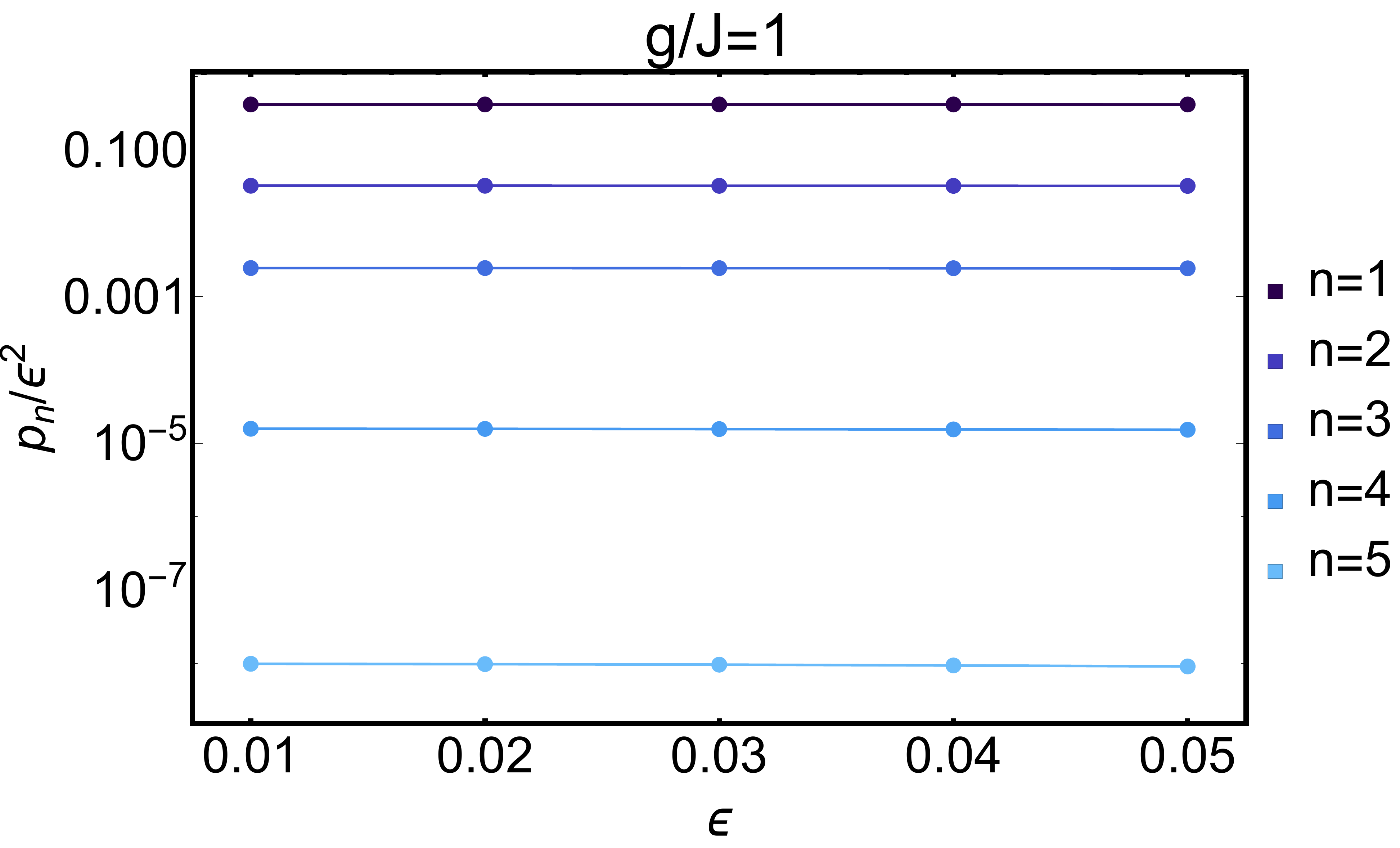}
\hspace{0.25cm}
\includegraphics[width=0.29\textwidth]{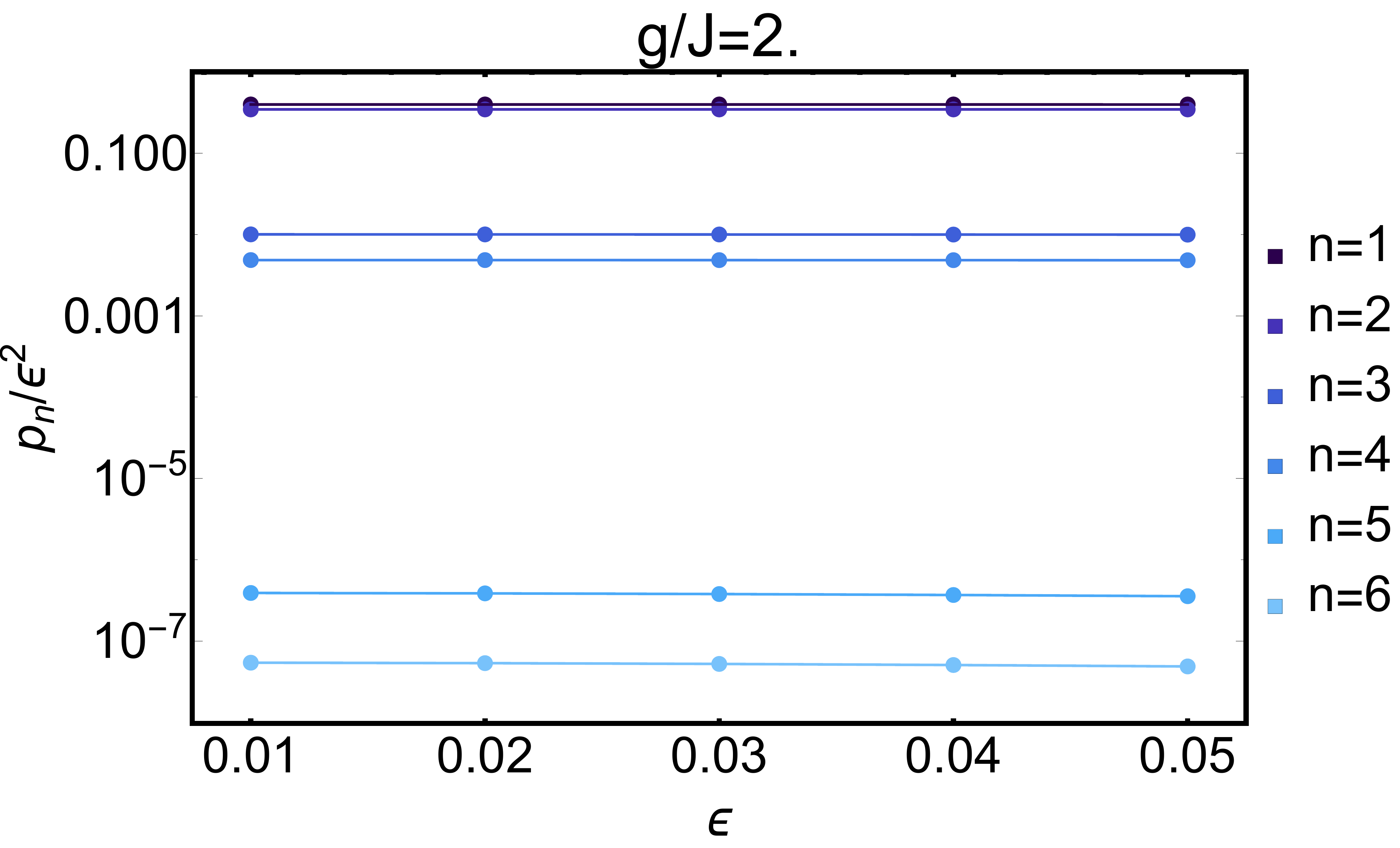}
\caption{Behavior of the single-particle entanglement spectrum of the IM of the transverse-field quantum Ising chain.
The largest eigenvalues $0\le p_n \le 1/2$ of the correlation matrix of a half-time interval originating from the IM wavefunction $\ket{\Psi}$ (see the text) are reported, rescaled by $\epsilon^2$, as a function of $\epsilon$, for three values of $g/J$ reported above the plots.
In all cases, the results indicate the scaling behavior $p_n \sim \epsilon^2 P_n$, where $P_n$ are $\epsilon$-independent quantities.
}
\label{figctl}
\end{figure*}

We are now in a position to understand the continuous-time limit in this solvable case. Setting $J=\epsilon$, $T=t/\epsilon$,
we find that the variable $J^2 T = \epsilon t$ flows to zero for any fixed value of the physical evolution time $t$. Since $P \sim \epsilon^2$ as $\epsilon\to0$, the von Neumann entropy of any extensive bipartition converges to zero in the continuous-time limit.
It is instructive to note that in this case the entanglement spectrum is composed of only two values, $P$ and $1-P$.
When $\epsilon$ is sufficiently small, one might be tempted to truncate the corresponding singular value (e.g., when performing a standard SVD compression). This seemingly harmless approximation, however, would produce incorrect results. Indeed, after such a truncation, the IM becomes a product state relative to the considered bipartition, say at the bond $(M,M+1)$: 
physically, this corresponds to ``refreshing the environment'', i.e., tracing it out after $M$ time steps and replacing it with another infinite-temperature environment before time step $M+1$.
Performing a similar truncation in other bonds, we end up with an IM corresponding to a ``noisy'' classical magnetic field along $\hat z$ switching between $\pm J$, whereas the actual exact IM in Eq.~\eqref{eq_IMg0} represents the influence of a constant field.
%
We thus conclude that in the  continuous-time limit, truncation of small Schmidt values might be dangerous, and may not lead to an accurate representation of system's dynamics. Furthermore, counter-intuitively, the mere value of temporal entropy is not representative of the extent of temporal correlations and of the computational resources needed to parametrize the IM.

We have further confirmed this picture by an analytical calculation of the IF in the integrable limit $h=0$~\cite{IntegrablePaper}.
Making use of the quadratic fermionic representation of the model, $S_t$ is reduced to the entropy of a Bardeen-Cooper-Schrieffer-like wavefunction on the Keldysh contour, due to its Gaussian form.
In this case, as shown in Fig.~\ref{figctl}, the scaling of $S_t$ with $\epsilon$ stems from an $\epsilon^2$ global scaling of the single-particle entanglement spectrum, $p_n \sim \epsilon^2 P_n$, which gives
\begin{multline}
S_{\tau,t} \equiv -\sum_n p_n \log p_n + (1-p_n) \log(1-p_n)  \\ \sim  \epsilon^2 [\log (1/\epsilon^2) + \hat S ],
\end{multline}
where $\hat S = -\sum_n P_n \log P_n + (1-P_n) \log(1-P_n) $ is a finite, $\epsilon$-independent quantity.
Once again, our tentative conclusion that all the singular values could be safely truncated as $\epsilon\to0$  would lead to an incorrect description of dynamics, as the resulting product-state form of the IM would be incompatible with the existence of finite temporal correlations in the continuous-time limit.

The behavior of the IM at $\epsilon\to 0$ in the solvable cases discussed above indicates that caution must be exercised when applying conventional compression techniques to influence functionals. Thus, in the future it would be beneficial to develop other numerical schemes that take into account the structural constraints of IMs in the continuous-time limit.



\section{Non-ergodic dynamics: discrete time crystals}\label{sec4}




\begin{figure*}[t]
\centering
\includegraphics[width=0.7\textwidth]{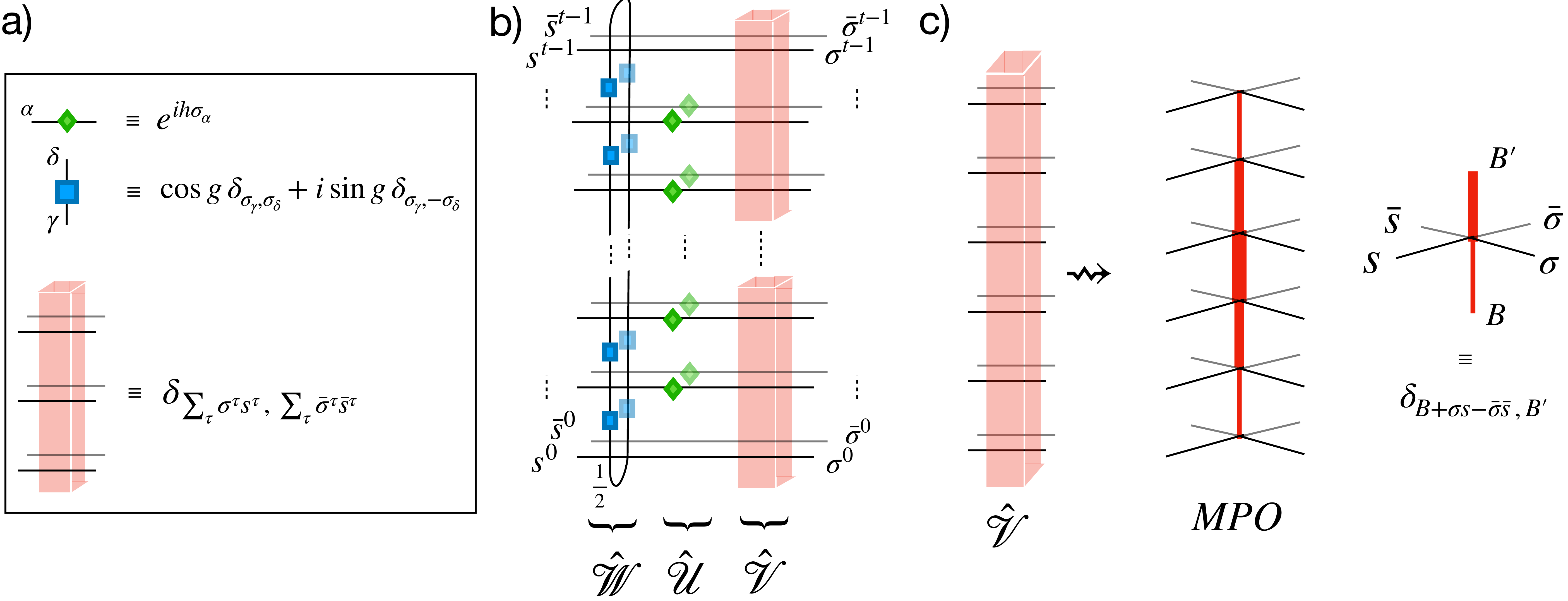}
\caption{Depiction of the tensor network corresponding to the discrete path
integral of the Floquet model with fully random $J$. Averaging over $J$ can be
implemented by an operator $\delta_{\sum s \sigma \sum{\bar{s}\bar\sigma}}$.
This operator can be implemented as an MPO by introducing a virtual index which
at each spin corresponds to the difference in partial sums up to this spin. The
condition can be enforced by only allowing partial sum values that are
consistent with it. The maximal bond dimension of the resulting operator is
$t$.}
\label{figtctn}
\end{figure*}

\begin{figure*}[t]
\centering
\includegraphics[width=0.7\textwidth]{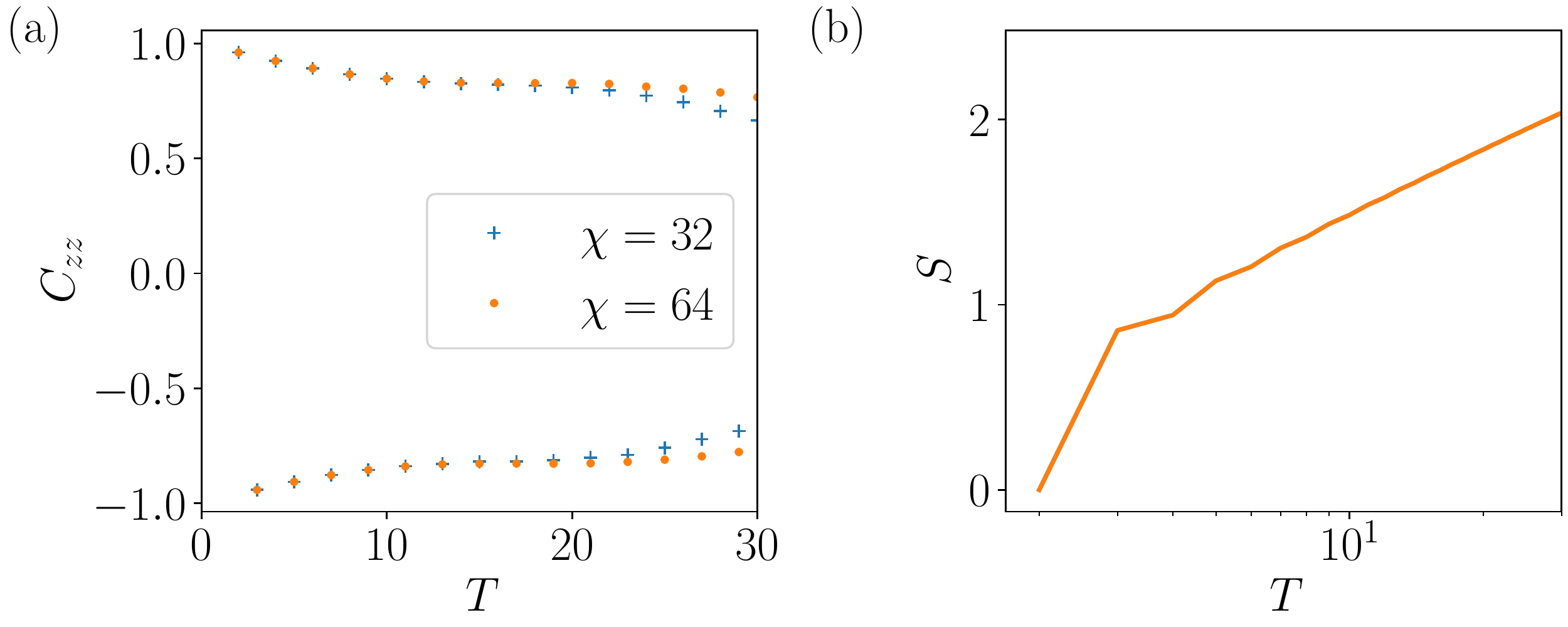}
\caption{Discrete time crystal simulated using influence functional approach.
We use a Floquet model in Eq.~\eqref{eq:DTC} with parameters $g=\frac{\pi}{2}-0.1,h=0.3$, and $J$
fully random. The initial state corresponds to an infinite temperature ensemble. {\it
Left:} $C_{zz}$ correlations show persistent oscillations. For larger time, the
amplitude of oscillations decreases for bond dimension $\chi=32$, which is attributed to the effects
of the compression. {\it Right:} TE grows logarithmically with evolution time. Two curves for different bond dimensions coincide.}
\label{fig_5}
\end{figure*}
The influence functional approach allows to treat the effects of spatially
uncorrelated randomness via exact ensemble averaging, denoted by $\mathbb{E}\big(
\cdot \big)$. As in the Schwinger-Keldysh path integral
formalism~\cite{Kamenevbook2011}, ensemble-averaged observables can be found
from  the ensemble-averaged IF, without the need of introducing replicas.
Provided randomness in different spatial points is uncorrelated (or short-range
correlated), and provided its distribution is spatially homogeneous, the
ensemble-averaged IF satisfies a self-consistency equation defined by the
transfer matrix $\mathbb{E}\big( \tilde{T} \big)$.
This applies to various kinds of randomness. Above we have already introduced
ensemble-averaging over statistical mixtures of initial states (in particular,
infinite-temperature density matrices). Random spatiotemporal noise can also be
easily incorporated; in Ref.~\cite{LeroseInfluence}, we showed that circuits with
fully random Haar-distributed kicks and fully random  diagonal interactions
behave as perfect dephasers (almost surely in the ensemble distribution in the
limit of large local Hilbert space).
Furthermore, in Ref.~\cite{Sonner20CharacterizingMBL}, the formalism was applied to systems with quenched disorder. In this case, the random variables are constant in time, and exact averaging over them introduces  non-local in time
effective interactions in the problem.

One of the most appealing aspects of the IF approach is that
the self-consistent IF fully encodes the ability of a quantum many-body
system to provide an efficient thermal bath. In particular, it should be
possible to classify universality classes of quantum dynamics based on the
structure of their IF, e.g., to distinguish between ergodicity vs non-ergodicity.
It is known that strongly disordered quantum lattice systems can break
ergodicity via the mechanism of many-body localization (MBL)~\cite{AbaninRMP}.
Reference~\cite{Sonner20CharacterizingMBL} began to explore the MBL transition
using the IF approach.
The model considered in Ref.~\cite{Sonner20CharacterizingMBL} is the kicked
Ising chain of Eq.~\eqref{eq:KIM} with random longitudinal fields $h_j$
uniformly distributed in $[0,2\pi)$. This model exhibits a Floquet-MBL transition
upon varying the relative strength of the transverse field and/or of
interactions~\cite{ZhangPRB16_FloquetMBL,Sonner20}.

Discrete time crystals (DTCs) represent a notable regime of
non-equilibrium quantum dynamics. DTC behavior in periodically driven quantum
lattice systems is associated with a robust spatiotemporal order that
spontaneously breaks the discrete time-translation symmetry. A DTC phase
requires a coherent periodic drive and an underlying physical mechanism protecting
the system from indefinite heating, such as MBL or prethermalization~\cite{PRBPrethermal2017,ElsePrethermal2017}. DTC response is robust against arbitrary (weak) perturbations of the
system interactions and of the driving protocol~\cite{Khemani16,Else16}.
Below we show that the IM approach provides a suitable tool for this phenomenon as well.
Indeed, the IM formalism naturally incorporates the necessary ingredients to
observe time-crystalline behavior, namely discrete time-translation symmetry,
thermodynamic limit, and disorder averaging.

DTC response can be observed, in particular, in a kicked Ising model
with random interactions and single-spin kicks close to perfect $\pi$-rotations
around the $\hat x$ axis~\cite{Khemani16,Else16}. We consider the model in Eq.~\eqref{eq:KIM}, introducing bond-dependent couplings $J_j$ uniformly distributed in $[0,2\pi]$, and setting $g=\pi/2-\varepsilon$:
\begin{equation}
\label{eq:DTC}
U=e^{-i (\pi/2 - \varepsilon) \sum_j \sigma_j^x} \, e^{-\sum_j J_j \sigma_j^z \sigma_{j+1}^z -h\sum_j \sigma_j^z}.
\end{equation}
For $\varepsilon=0$, the kick commutes with the interaction term: every initial product state in the $z$-basis undergoes perfect flips of all the spins each period, and thus returns to the initial configuration every other driving period, exhibiting a trivial DTC dynamics.
When $\varepsilon$ is non-zero, but sufficiently small, the subharmonic DTC response of the system survives many-body quantum fluctuations. The robustness of the DTC is made possible by MBL, which prevents the domain-wall excitations generated by the periodic quenches from spreading across the chain and melting the spatiotemporally ordered pattern, as it would happen in the absence of disorder.

To analyze DTC behavior using the IF formalism, we first note that averaging over $J$ in the dual transfer matrix can be done exactly:
\begin{equation}
\label{eq_exactdisav}
    \int \frac{\mathrm{d}J}{2\pi}  \prod_{\tau} \exp\left[iJ(\sigma^\tau s^\tau -\bar{\sigma}^\tau\bar{s}^\tau )  \right] = \delta_{\sum _{\tau}{\sigma^\tau s^\tau},\sum_{\tau} {\bar\sigma^\tau \bar s^\tau}} \, .
\end{equation}
The resulting operator can be interpreted as an adjacency matrix, which allows a
transition between trajectories $\sigma_\tau,\bar\sigma_\tau$ and $s_\tau,\bar
s_\tau$ as long as the total number of spins flipped on the forward trajectory
is equal to the total number of spins flipped on the backward trajectory. We can
express this operator as a MPO as depicted in Fig.~\ref{figtctn}: Each tensor in
the MPO is given by $\delta_{B+\sigma^\tau s^\tau-\bar{\sigma}^\tau \bar s
^\tau, B'}$ {where the virtual indices $B,B'$ carry the running sums of flipped spins on the forward minus those on the
backward trajectory, such that the 
averaging condition \eqref{eq_exactdisav} is globally enforced}. This procedure yields an MPO with bond
dimension $t$. Since bond dimension of the operator is relatively large, we
cannot apply the MPO to the MPS and compress the result. Instead, the
compression needs to happen during the application of the operator \cite{zipup}.

To detect subharmonic response, we computed the (disorder-averaged) IM using an MPO-based algorithm, and used it to calculate the dynamical correlation function $\mathcal{C}_{zz}(T)$. The model (\ref{eq:DTC}), similar to closely related models considered in Refs.~\cite{Khemani16,Else16}, exhibits a DTC phase at sufficiently small detuning from a perfect $\pi$-rotation, $\varepsilon$. This is evident in Fig.~\ref{fig_5}, which illustrates $\mathcal{C}_{zz}(T)$ for $\varepsilon=0.1, h=0.3$. The correlator exhibits persistent oscillations with an amplitude that stays sizeable at the times simulated. Note the slight decay at times $T>20$ for the smaller bond dimension; we attribute it to the errors introduced by compression to an MPS form. Indeed, increasing the bond dimension $\chi$ restores the stable oscillatory behavior for longer times.
We have further observed that for $\varepsilon \gtrsim 0.3 $ the behavior changes qualitatively, and the correlation function decays at short, $\chi$-independent time scales, suggesting that the system enters a thermalizing phase.

To assess the efficiency of the method, we have also analyzed the scaling of TE entropy with $T$ (right panel of Fig.~\ref{fig_5}, converged with respect to $\chi$), finding it to be approximately logarithmic. To understand the origin of this behavior, let us consider the trivial DTC point $\varepsilon=0$, i.e., $g=\pi/2$. In this case quantum fluctuations are completely suppressed, and, analogously to Eq. (11), the exact IM for spin $j$ reads

\begin{equation}
 I[\sigma_\tau,\bar\sigma_\tau] = \cos\big[ J_j \sum_\tau (-)^\tau (\sigma_\tau - \bar\sigma_\tau)\big].
\end{equation}
Taking the exact average over the random coupling $J_j$, we obtain
\begin{equation}
 \mathbb{E} \big( I[\sigma_\tau,\bar\sigma_\tau] \big) = \delta_{ \sum_\tau (-)^\tau \sigma_\tau \, , \,  \sum_\tau (-)^\tau  \bar\sigma_\tau}.
\end{equation}
Viewed as a wavefunction in the folded multi-time Hilbert space, similarly to what done in Eq. (12), we find that
\begin{equation}
\mathbb{E} \big( I[\sigma_\tau,\bar\sigma_\tau] \big) \mapsto
\Psi =
\bigg( \bigotimes_\tau
\frac {1}{\sqrt{2}}
\begin{pmatrix} 1 \\ 1 \end{pmatrix}_{cl,\tau}
\bigg)
\otimes
\bigg\lvert\frac{T}{2},0 \bigg\rangle_{q}
\end{equation}
where $\ket{\frac{T}{2},0}_{q}$ represents the Dicke state of  $T$ spins-$1/2$ with \textit{staggered} collective spin magnitude $S=\frac T 2$ and magnetization $S_z=0$. The entanglement entropy of this wavefunction scales as $\frac 1 2 \log T$.
This form of temporal entanglement stems from the effective long-range interactions in time arising from exact disorder averaging.
(We observe that a similar conclusion applies to the case $g=0$, where the collective spin magnetization is uniform rather than staggered.)

We finally note that Ref.~\cite{Sonner20CharacterizingMBL}, which considered Floquet-MBL phase in a random-$h$ kicked Ising model, found that the scaling of TE entropy was strongly sublinear in the MBL phase, qualitatively similar to what is reported above.
However, we note that there a variational DMRG approach to computing the IM was used, while here we employed iterative algorithm. In the future, it would be interesting to compare the performance of the two approaches to solving self-consistency equation for different models.

\section{Summary and outlook}\label{sec5}

%

In this paper, we characterized dynamics of a many-body system via its influence functional -- an object commonly used for describing dynamics of open quantum systems. IF provides a tool for quantifying the ability (or lack thereof) of a many-body systems to serve as a thermal bath for its parts~\cite{LeroseInfluence,Sonner20CharacterizingMBL}.
One of our goals was to illustrate the versatility of this new approach, by applying it to describe dynamics in very different regimes and setups. In particular, we studied thermalization in chaotic Floquet and Hamiltonian models, as well as its absence, accompanying the DTC response, in disordered systems.

A central role in our analysis was played by the temporal entanglement of the IF ``wavefunction", which is an analogue of the conventional real-space entanglement of many-body wavefunctions. Similarly to the case of ground states in many-body systems, temporal entanglement serves as  a measure of the computational complexity of parametrizing the IF by a MPS. As discussed above, in several physical situations of interest, temporal entanglement remains low, providing new corners where many-body dynamics can be efficiently simulated (and, in some cases even solvable). {Furthermore, we expect that other measures of temporal entanglement of IF could provide a witness of universal aspects of dynamics, for example, distinguishing between ergodic and non-ergodic dynamics. }

Looking forward, the IF formalism appears suitable for studying entanglement transitions in hybrid quantum circuits composed of unitary gates and projective measurements or local dissipation~\cite{NahumPRX19_MeasurementInduced,Fisher19_MeasurementDriven}: indeed, both randomness and dissipation can be naturally incorporated in this approach. We expect that the character of the asymptotic state of such systems is encoded in the structure of the self-consistent, ensemble-averaged influence matrix.
On a more practical level, the development of efficient numerical schemes to approximate the IF of generic quantum many-body systems, while challenging, is a promising direction for future research, which may provide new insights into several long-standing problems in quantum dynamics, including the nature of the MBL-thermal transition~\cite{PotterPRX15,VoskPRX15,ZhangPRB16,DumitrescuPRL17,ThieryPRL18,GoremykinaPRL19}.

Finally, we hope that the approach outlined above will stimulate new fruitful connections between theory of open quantum systems and quantum many-body dynamics~\cite{TEMPO,Cygorek21,Chan21}. For example,  one can envision that standard concepts of the former -- such as memory kernels~\cite{MakriMakarov94} and non-Markovianity measures~\cite{BreuerRMPNonMarkovian} --  may be turned into new tools for the latter.  Progress in this direction may enable classification of universal basins of attraction in quantum many-body dynamics. In addition, ideas and tools from quantum many-body physics, including tensor-networks, could widen the scope of open quantum systems theory, for example, allowing one to describe the effects of truly interacting, quantum chaotic environments.

\section{Acknowledgments}

This work was supported by the Swiss National Science Foundation and by the European Research Council (ERC) under the European Union's Horizon 2020 research and innovation programme (grant agreement No. 864597). 
We thank Lorenzo Piroli and Soonwon Choi for useful discussions. Computations were performed at the University of Geneva on the ``Baobab'' and
``Yggdrasil'' HPC clusters.

\bibliography{mbl}

\end{document}